\documentclass[preprint,12pt]{revtex4}
\usepackage{graphicx,epsfig}
\usepackage{caption,subfigure}
\begin{document}
\title{Dynamics of atomic and molecular solitary waves in atom-molecular hybrid Bose-Einstein condensates coupled by Magnetic Feshbach Resonance: Role of induced decays of Feshbach Molecules}
\author {Krishna Rai Dastidar$^{2,b}$ and Deb Shankar Ray$^{2,a}$
\\$^2$ Indian Association for the Cultivation of Science,
\\Kolkata 700032, India
\\$^b$krishna.rai.dastidar@gmail.com
\\ $^a$pcdsr@iacs.res.in}
\date{}
\begin{abstract}
 Dynamics of atomic and molecular bright solitons in a hybrid atom-molecular BEC (Bose Einstein Condensate) system of $^{85}$Rb coupled through Magnetic Feshbach Resonance (MFR) has been investigated. By solving the time independent atom-molecular coupled equations, the initial atomic and molecular wavefunctions were obtained and the dynamics of the initial atomic and molecular waves in a spherical one-dimensional trap were studied by solving the time-dependent atom-molecular coupled equations. During evolution two types of induced or stimulated decays of the feshbach molecules were switched on. These two types of stimulated decays of the feshbach molecules (i) to the two-atom continuum and (ii) to the bound level of the lowest hyperfine state  of the molecule were induced by laser/RF (Radio Frequency) fields. Hence the strength of these two induced decays can be controlled by varying the laser/RF field  parameters e.g. intensity, detuning etc. It is found that depending on the relative strength of these two types of stimulated or induced decays initial atomic and molecular waves assume solitonic  nature as bright solitons during evolution and the stability of these solitonic waves can be achieved by controlling the relative strength of induced decays in two different channels. It is shown that these two induced decays lead to the formation of stable atomic and molecular solitons by suppressing the initial oscillations and instability in the atom-molecular coupled system of $^{85}$Rb atoms.


\end{abstract}
\maketitle
\section{Introduction}
	
The macroscopic behavior of an ultracold dilute atomic gas can be tuned by controlling the interaction between two atoms
due to low energy binary collisions. Most fascinating feature of the macroscopic behavior is the conversion of ultracold atoms
into molecules which can be achieved by tuning the ultracold atomic system through Feshbach Resonance (FR).  Extensive theoretical and experimental studies on hybrid ultracold atom-molecular coupled BEC (Bose Einstein Condensate) establish that
 substantial amount of molecular formation in atomic BEC is feasible through magnetic as well as optical
(single or two photon coupling) Feshbach resonance [1-24]. In this system
atom-molecular coupling through Feshbach resonance ( magnetic or optical) is responsible for atom to molecular conversion and vice-versa.

In ultracold atom-molecular hybrid system coupled through optical or magnetic Feshbach Resonance, the existence of the coherence between atoms and molecules was
demonstrated by studying the evolution of the atomic and molecular densities.
In presence of spherical and asymmetrical trap it is found that
in a coupled system both the atomic and molecular density oscillate with time and this oscillation is out of phase[4,20]. This is a
signature of induced coherence in the system due to atom-molecular coupling. In optical feshbach resonance two ground state atoms sitting
in the dissociation limit of the ground state molecule are photo-associated to form molecules in the excited state whereas in magnetic feshbach resonance ground state atoms are transferred to a bound molecular state by hyperfine interaction induced by swaping magnetic field. Recently it has been demonstrated that magnetic feshbach resonance can be achieved to form molecules in the excited electronic state by swaping the magnetic field in ultracold excited atoms [8].

Although the generation of ultracold molecules are feasible by magnetoassociation or photoassociation of ultracold atoms through magnetic or optical  Feshbach resonance respectively, the efficiency for atom to molecule conversion is moderate. Attempts have been made to increase the efficiency of conversion and high efficiency has been achieved by the method of Stimulated Raman Adiabatic Passage (STIRAP) [25] and magnetoassociation followed by
adiabatic passage (using a laser field)[26]. Moreover in Magnetic Feshbach Resonance (MFR) a radio frequency (RF) field has been used to induce adibatic passage from two atom continuum state to different hyperfine bound states of Feshbach molecules [27]. Usually both the atomic and molecular densities oscillate out of phase with time due to coherence induced by atom-molecular coupling through Feshbach Resonances (FR). However for the generation of solitons with constant peak density inclusion of a damping is necessary [28]. Once the Feshbach molecules are formed in the upper hyperfine states by magnetoassociation, there is a possibility that it can decay via two channels (i) to the two-atom continuum and (ii) to the bound levels of the lowest hyperfine state of the molecules. Usually in case of MFR these two spontaneous decays are considered to be small. But these two decay channels can be stimulated by RF or laser fields depending on the energy difference between two hyperfine states. Stimulated or induced decay of the Feshbach molecules to the two-atom continuum may replenish ultracold two-atom states and the induced decay to the bound levels of the lowest hyperfine state
may facilitate formation of cold molecules depending on the strength of these two stimulated decays. Since these two decays are stimulated by RF or laser fields the relative strength of these two decays can be controlled by controlling the field intensity, detuning etc. In this paper we have investigated a scheme for the generation of stable atomic and molecular solitons  from ultracold atoms and molecules of $^{85}Rb$ coupled through MFR by stimulating decays of Feshbach molecules in two different channels and controlling the relative strength of these two induced decays. In recent experiment [27] modification of MFR has been achieved by coupling the two-atom continuum to the bound levels of the Feshbach molecule using RF fields. In the present work we have considered two RF fields to couple bound hyperfine levels of upper hyperfine state to the two-atom continuum and to the lowest bound level of the lower hyperfine state respectively to control the generation of stable solitary waves (Fig. 1). In view of recent experiment [27] study of the present modified MFR system may be experimentally feasible.

To generate solitonic waves in ultracold atomic system, usually a cigar-shaped trap where the axial component of the trap is
released keeping the spherical traping unchanged. In this configuration solitonic waves are formed and its  solitonic nature persists for some time.
The nature of the solitonic waves in atomic BEC is governed by the strength and the sign of interactions arising from atom-atom
scattering. Atom-atom interaction is proportional to the s-wave scattering length and hence the strength and the sign of
the interaction will depend on the nature of the scattering length.  In a dilute weakly interacting
ultracold bose gas different type of solitons are formed due to the interplay between positive dispersion and positive or negative atom-atom interactions.
In atomic BECs the formation of different type of
 solitons in different atom-atom interaction regime have been
 observed and also studied theoretically [28-42].

In atom-molecular hybrid BEC system coupled through magnetic (magnetoassociation) and optical (photoassociation) Feshbach resonance, the formation of atomic and molecular solitons  and its stability have been studied theoretically [43-50].
 In this system in addition to the positive dispersion and interaction arising from atom-atom scattering other interactions due to atom-molecule
and molecule-molecule scattering (which are proportional to their respective scattering lengths) enter into the picture determining the strength and the sign of net interaction due to
boson-boson scattering.
In addition to the boson-boson interactions the presence of atom-molecular coupling  which is attractive in nature will affect
the nature of soliton formation (both atomic and molecular) and its stability.  Recently this aspect of the  effect of atom-molecular coupling on the formation of stable atomic and molecular solitons in atom-molecular BEC system coupled through Raman Photoassociation (PA)  has been explored in detail [50].
In case of two photon Raman photoassociation two laser fields are applied to couple  the excited bound state of the Feshbach molecule to the two atom continuum and  to the lowest bound state of the molecule respectively. Presence of these two laser fields leads to induced decay of the excited molecule to the two-atom continuum  and to the bound vibrational levels of the ground state. Previously it has been shown that in case of two photon Raman photoassociation these two induced decays  play a crucial role to determine the stability of solitonic waves [50].  However in case of MFR these two decays can be stimulated by applying RF fields. By channelizing the decay of the Feshbach molecules in two different paths to facilitate formation of ultracold two atoms or formation of cold molecules in the ground state generation of stable atomic and molecular solitons can be controlled. It will be shown that these two induced decays play a crucial role to damp the irregular oscillations and instability in  amplitude of evolving matter waves leading to generation of stable bright solitonic waves.

In this work we have used experimentally realized parameters for MFR in $^{85}Rb$ BEC [4] and introduced stimulated
decays of the Feshbach molecules ( to the bound and to the two-atom continuum of the ground state) which can be induced by RF fields.

\section{Theory}
Our model is an ultracold hybrid atom-molecular system coupled through magnetic Feshbach Resonance where decay of the Feshbach molecule from the bound level of the upper hyperfinr state to the two-atom continuum of the lower hypefine state and to lowest bound level of the lower hyperfine state has been induced by applying electromagnetic fields. Schematic diagram of this model system has been shown in Fig.1. For this coupled system Gross-Pitaevskii (GP) like
equations can be obtained for the atomic and molecular condensate
wave functions.  By considering
the induced decays from the molecular excited states,
the GP like equations of motion for the coupled (MFR) condensate system
become
$$i\hbar {\partial \psi_a\over \partial t}=[-{\hbar^2\nabla^2\over
2m}+U_a(\vec r)+\lambda_a
|\psi_a|^2+\lambda_{am}|\psi_m|^2]\psi_a+\chi
\psi_m\psi_a^*-i\hbar\alpha\psi_a-$$\begin{eqnarray}i\hbar\Gamma_1|\psi_a|^2\psi_a\end{eqnarray}
and\begin{eqnarray}i\hbar {\partial \psi_m\over \partial
t}=[-{\hbar^2\nabla^2\over 4m}+U_m(\vec
r)+\epsilon+\lambda_m|\psi_m|^2
+\lambda_{am}|\psi_a|^2]\psi_m+{\chi\over
2}\psi_a^2-i\hbar\Gamma_2\psi_m\end{eqnarray}
where $m$ is the atomic mass, $\lambda_a$, $\lambda_m$,
$\lambda_{am}$ are the atom-atom, molecule-molecule and
atom-molecule interaction strengths respectively. For spherically symmetric trap the external trap potentials for atoms and
molecules  $U_a(\vec r)$ and
$U_m(\vec r)$ are given by $U_a(r)={1\over 2}m\omega^2r^2$ and
$U_m(r)=m\omega^2r^2$, where $\omega$ is the angular frequency and
$r$ is the radial distance. Within Bogoliubov mean-field theory
\cite{Bogo} atom-atom interaction strength
$\lambda_{a}={4\pi\hbar^2a/m}$, where $a$ is the s-wave scattering
length for atom-atom interaction. We assume here for simplicity
$\lambda_m=\lambda_{am}=\lambda_a$.  Here $\Gamma_{1,2}$ are the induced decay rates of Feshbach molecules
to the two-atom continuum and to the bound levels of the ground state respectively and $\alpha$
  is the decay rate of the excited atoms lying in the continuum of upper hyperfine state. The rate of
spontaneous emission from excited molecular state and that from the excited atomic state have been included in
$\Gamma_{1,2}$ and $\alpha$ respectively. For magnetic Feshbach resonance the  induced decay rates $\Gamma_{1,2}$
can be induced or stimulated by applying RF or laser fields to couple the upper bound hyperfine level of the Feshbach molecule
to the two-atom continuum (by L1 as shown in Fig.1) and to the lowest bound level (by L2 as shown in Fig.1) of the ground hyperfine state respectively.
The atom-molecule coupling strength (arises due to Feshbach
resonance) $\chi$ can be given in terms of resonance width
($\Delta B$), the difference between magnetic moments of a
molecule and a free atom pair ($\Delta \tilde \mu$) and background
scattering length of atoms far from a Feshbach resonance
($a_{bg}$): $\chi= \sqrt {8\pi\hbar^2a_{bg}\Delta {\tilde \mu}
\Delta B/m}$.
For spherically symmetric trap two 3D coupled equations (1) and (2) reduce to effectively one-dimensional form form as a function of r (radial distance) where $\psi_{a,m}(\vec{r},t)$
is given as $\phi_{a,m}(r,t)/r$ and the boundary conditions for solving these two equations are $\phi_{a,m}(r,t)$ $\rightarrow$ $0$ as $r$ $\rightarrow$ $0$ and $\infty$ respectively [20-21].

To solve coupled equations for spherically symmetric trap we have used initial atomic and molecular wave functions which were obtained by solving the time-independent coupled GP like equations [50] for the ultracold atom-molecular system coupled through magnetic feshbach resonance by putting $\Gamma_1$=$\Gamma_2$=$\alpha$=0. Then to study the evolution of these initial atomic and molecular waves we have solved the time-dependent coupled equations (GP-like) by switching on the decays $\Gamma_1$, $\Gamma_2$ and $\alpha$.

 Solution of these two dynamical equations 
 for atomic and molecular wave functions gives the dynamics
of the coupled system for spherically symmetric trap in GP approach. For solving the equations we have considered the one dimensional model [37,48,50] and written the equations in terms of  dimensionless quantities x and $\tau$ such that $r=x  a_{ho}$ and $t=\tau /\omega$  where $a_{ho}=\sqrt{\hbar/{m \omega}}$ [20].
To solve the one dimensional coupled
equations we use the Crank-Nicholoson scheme to discretize the
equations and then the tridiagonal set of equations are solved by
Gauss elimination and back substitution method.
 The details of the
numerical method has been given in [20-23].

Previously we have demonstrated [20] the evolution of atom-molecular
coherence in $^{85}$Rb condensate in presence of magnetic Feshbach
resonance for which this type of coherence has been observed
experimentally [4] by solving the
coupled GP equations for magnetic Feshbach resonance (equations (1) and (2) by putting $\Gamma_1=\Gamma_2=0$.
We consider $^{85}$Rb condensate in a spherically
symmetric trap with frequency $\omega/2\pi$= 12.69 Hz.
 Other parameters used are N=17100, atomic mass $m= 1.4112568490 \times {10^{-25}} kg$, $a$= 570 $a_0$, $a_{bg}$= -450$a_0$ (where
$a_0$ is the Bohr radius), $\Delta B$=11 Gauss, width of resonance $\Delta$= 11 Gauss, decay time for the excited atoms is 91
$\mu$s, $\Delta \tilde \mu$= -2.23 $\mu_B$
(where $\mu_B$ is the Bohr magneton)and the detunning $\epsilon= 200 kHz $ at B= 159.84 $G$. Hence
atom-molecular coupling strength $\chi = 3107.377 \times 10^{-7}  m^{3/2}$ $Hz$, atomic induced decay rate $\alpha = 2.1739 \times 10^{4} Hz$. For this calculation we have used the same parameters for $^{85}Rb$ atoms mentioned above. However  the  molecular induced decay rates $\Gamma_1$ and $\Gamma_2$ have been  varied for designing of stable atomic and molecular solitons, keeping other parameters fixed.

\section {\textbf{Results and Discussions}}

In this paper we have studied the generation and stability of atomic and molecular solitonic matter waves (bright solitonic nature) in a hybrid atom-molecular BEC system of $^{85}$Rb atoms coupled
through magnetic feshbach resonance by inducing the decay of the Feshbach molecule in the hyperfine state (i) to the two-atom continuum and (ii) to the lowest bound levels of the lower hyperfine state.

 In the present work the evolution of density profiles of atomic and molecular matter waves have been studied with the trap on for 3.75 secs considering the stimulated decays of the excited Feshbach molecules  ($\Gamma_1$ and $\Gamma_2$ ). To study the dependence of the evolution of atomic and molecular density profiles on the induced decay of the Feshbach molecules in two channels we have varied the strength of induced decays $\Gamma_1$ and $\Gamma_2$ keeping other parameters (e.g. N, $\epsilon$, $\alpha$ etc) fixed. It is to be mentioned that the decay of excited atom ($\alpha$) is very small compared to induced decays of Feshbach molecules ($\Gamma_1$ and $\Gamma_2$) considered here. It can be shown that the results presented here remain unchanged if $\alpha$ is neglected. Here $\Gamma_1$ has been varied from $1.629 \times 10^{-23}$ to $1.629\times 10^{-5}$ (in the units of $m^3 Hz$)  for each value of $\Gamma_2$, ranging from $304.4 \times 10^8 Hz$ to $304.4\times 10^{14} Hz$. This system is similar to the 3-level $\Lambda$ system where two arms of the $\Lambda$ scheme are provided by two channels of induced decays of the Feshbach molecules residing in the  bound levels of the upper hyperfine state. Hence these two induced decay channels compete with each other and lead to formation of stable atomic and molecular solitons depending on their relative strengths.   In this study we have investigated the stability region of both the atomic and molecular bright solitons as a function of $\Gamma_1$ and $\Gamma_2$. In table I we have shown stability of atomic and molecular solitons for different combinations of $\Gamma_1$ and $\Gamma_2$. It shows that when the strength of both the induced decays are small both the atomic and molecular solitons are unstable. This is due to the fact that the damping effect due to these two induced decays is not sufficient to suppress the initial oscillations/instability in the evolving atomic and molecular waves.  But as the strength of $\Gamma_1$ is increased keeping $\Gamma_2$ fixed ($304.4\times 10^{8} Hz$), atomic solitons become stable for high value of $\Gamma_1$ at $1.629\times 10^{-9}$ $m^3 Hz$. This means that when the damping due to $\Gamma_1$ is sufficiently strong stable atomic solitons are generated even for small value of $\Gamma_2$. At this value of $\Gamma_1$ molecular solitons become stable only when $\Gamma_2$ is  increased to the value $304.4\times 10^{12}$Hz.  This feature gets saturated if   $\Gamma_1$ is further increased by four orders of magnitude. For this value of $\Gamma_2$ ($304.4\times 10^{12} $Hz) stable atomic and molecular solitons are also generated simultaneously if the value of $\Gamma_1$ is further decreased to $(1.629\times 10^{-13}$ $m^3 Hz$) (Fig. 2). However for much smaller value of  $\Gamma_1=1.629\times 10^{-15}$ $m^3 Hz$, damping due to $\Gamma_2$ needs to be sufficiently large ($304.4\times 10^{14} $Hz) for the generation of both stable atomic and molecular solitons and this feature  persists if $\Gamma_1$ is decreased to lower value $1.629\times 10^{-17}$ $m^3 Hz$ (Fig. 3). For this highest value of $\Gamma_2$ ($304.4\times 10^{14}$ Hz) considered here if the value of $\Gamma_1$ is further decreased to $\Gamma_1=1.629\times 10^{-19}$, $\Gamma_1=1.629\times 10^{-21}$ and $\Gamma_1=1.629\times 10^{-23}$ $m^3 Hz$ the atomic soliton becomes quasistable, molecular soliton remaining stable. By quasistable it is meant that stable solitons with fine regular oscillations in amplitude are obtained as shown in Fig. 4 for $\Gamma_1=1.629\times 10^{-21}$. However this oscillation in amplitude is suppressed if the strength of  $\Gamma_1$ is increased as shown in Fig. 3. This type of solitons with oscillating amplitude have  been obtained in atom-atom coupled BEC system [52]. In figures (2-4) the evolution of atomic and molecular density profiles as a function of time has been shown in part (a) and (b) respectively.

 Therefore
the generation of stable bright atomic and molecular solitons from the atom-molecular hybrid BEC system coupled through MFR is feasible by controlling the relative strength of decay of the feshbach molecules to the two-atom continuum or to the bound level of the ground state and these two decays can be induced and controlled by RF or laser fields as mentioned before. This is due to the fact that these two induced decays for chosen optimum values can dominate over the existing oscillations or instability in the evolution of matter waves in $^{85}$Rb Bose Einstein condensates. Simultaneous generation of stable bright atomic and molecular solitons for a combination of optimum values of $\Gamma_1$ and $\Gamma_2$ can be attributed to the interplay between these two stimulated decay channels.

\begin{table}[h]
\caption{Stability of Bright Atomic and Molecular Solitons as a function of $\Gamma_1$ and $\Gamma_2$ } 
\centering 
\begin{tabular}{c rrrrrrrrrrr} 
\hline\hline 
$\Gamma_2$($304.4 Hz$) & $ 10^8$ && $ 10^{10}$ && $ 10^{12}$ &&
  $ 10^{14}$ \\ [0.5ex]
\hline 
$\Gamma_1$($1.629  m^3 Hz$)& Atom & Molecule & Atom & Molecule & Atom& Molecule & Atom &
Molecule \\ 
\hline
$  10^{-23}$ & Un & Un & Un & Un & Un& Un & Qs & St\\
$  10^{-21}$ & Un & Un & Un & Un & Un& Un & Qs & St\\
$ 10^{-19}$ & Un & Un & Un & Un & Un& Un & Qs & St\\
$ 10^{-17}$ & Un & Un & Un & Un & Un& Un & St & St\\
$ 10^{-15}$ & Un & Un & Un & Un & Un& Un & St & St\\
$ 10^{-13}$ & Un & Un & Un & Un & St& St & St & St\\
$10^{-11}$ & Un & Un & Un & Un & St& St & St & St\\
$ 10^{-9}$ & St & Un & St & Un & St& St & St & St\\
$10^{-7}$ & St & Un & St & Un & St& St & St & St\\
$ 10^{-5}$ & St & Un & St & Un & St& St & St & St\\
\hline 
$Un=Unstable;$ & $Qs=Quasistable;$ & $St=Stable$
\end{tabular}
\label{tab:hresult}
\end{table}
-------------------------

\section{Conclusion}
We have investigated the generation of stable atomic and molecular solitons in a hybrid atom-molecular BEC system of $^{85}Rb$ atoms coupled through magnetic feshbach resonance by introducing the stimulated decays of excited Feshbach molecules induced by RF/Laser fields. Two channels of decay of Feshbach molecule in the upper bound hyperfine level is induced by coupling the upper bound level to the two-atom continuum and to the lower bound hyperfine level of the ground state by RF fields.  To study the evolution of initial atomic and molecular matter waves  time-dependent coupled GP-like  equations for atoms and molecules have been solved by switching on the two induced decays of Feshbach molecules. The initial atomic and molecular waves were obtained by solving the time-independent GP like coupled equations neglecting all the decays. Initial atomic and molecular waves take the shape of bright solitonic waves during evolution depending on the choice of induced decay rates of the feshbach molecules. We have considered the other system parameters for $^{85}Rb$ atoms which have been experimentally and theoretically used to study the coherence of atomic and molecular waves in this coupled system. Evolving density profiles of atoms and molecules in effective one-dimensional trap (with trap on for ~3.75 secs) have been studied as a function of two induced decays ($\Gamma_1$ and $\Gamma_2$) to show the stability region  of the generated atomic and molecular solitons.    When the strength of both the induced decays is small solitons are unstable. But stable bright atomic or molecular solitons can be generated if the strength of $\Gamma_1$ or $\Gamma_2$ is increased to a high value irrespective of the strength of the other decay channels ($\Gamma_2$ or $\Gamma_1$) respectively.  This is due to the fact that when the damping due to induced decays is sufficient to suppress the initial oscillations/instability in atomic and molecular waves stable solitonic waves can be generated. It is found that for large value of $\Gamma_2$ but very small value of $\Gamma_1$ atomic solitons with small regular oscillations in amplitude are generated and this oscillation is suppressed with increase in the strength of $\Gamma_1$. However generation of stable bright solitons for both the atomic and molecular waves is feasible for a combination of optimum values of $\Gamma_1$ and $\Gamma_2$.  The interplay between these two induced (by RF fields) decay channels ($\Gamma_1$ and $\Gamma_2$) effectively controls the stability of both the atomic and molecular solitons leading to generation of stable bright atomic and molecular solitons simultaneously in the hybrid atom-molecular coupled (MFR) BEC of $^{85}Rb$.

\section {Acknowledgement}
We Thank  Department of Science and Technology, Govt. of India, for financial support under grant number SR/S2/LOP-13/2009.
\vfill\eject
\begin{figure}[!h]
\includegraphics[width=3.5in,height=3.5in,angle=0]{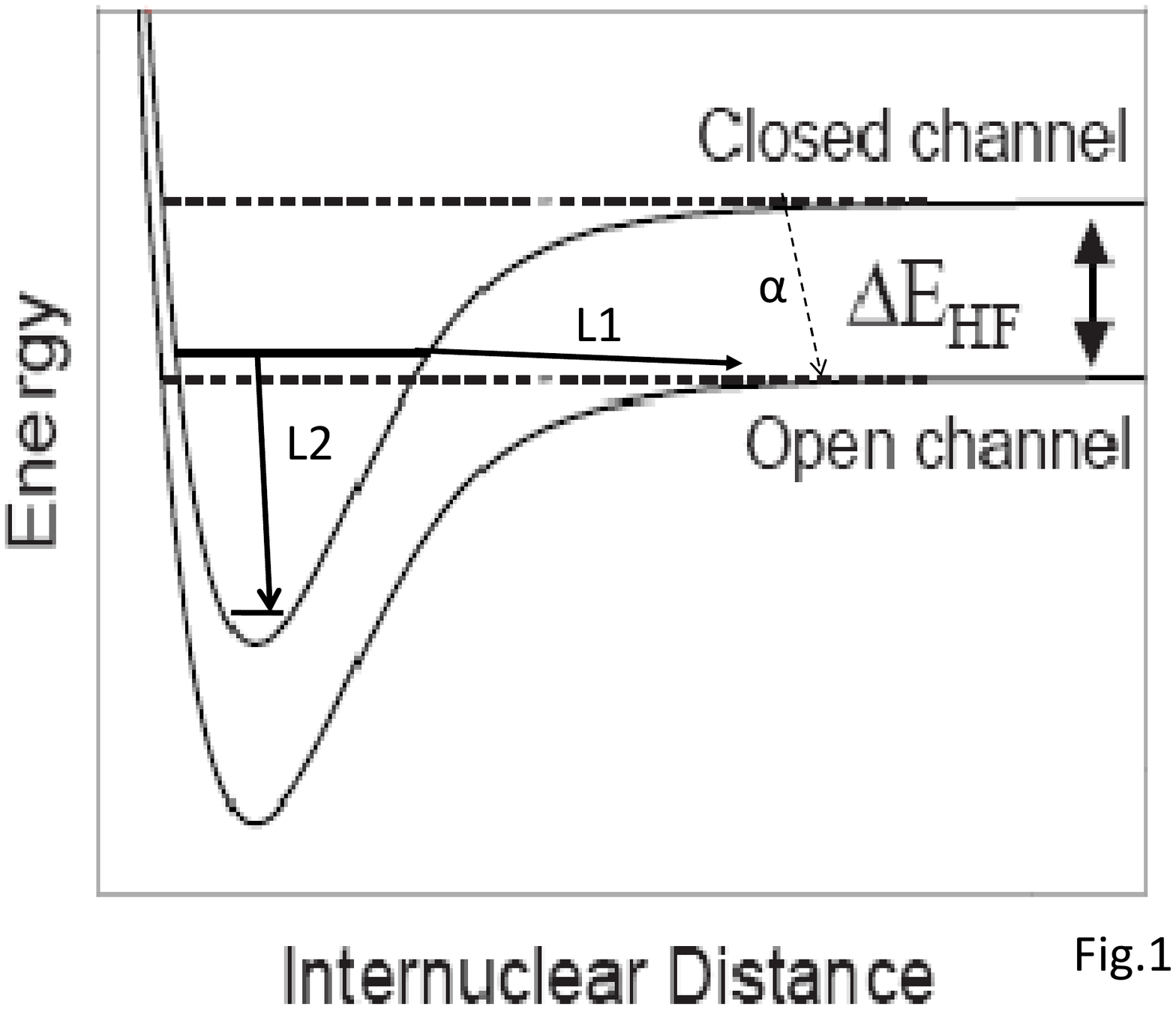}
\caption {Fig 1. Schematic diagram for modified Magenetic Feshbach Resonance (MFR) in atomic Bose Einstein Condensate (BEC) considered here. Decays of the Feshbach molecule formed in the bound level of the upper hyperfine state (closed Channel) to the two atom continuum of the lower hyperfine state (open channel) and to the lowest bound level of the lower hyperfine state have been induced by applying RF/Laser fields L1 and L2 respectively.}
\end{figure}
\begin{figure}[!h]
\includegraphics[width=3.5in,height=3.5in,angle=0]{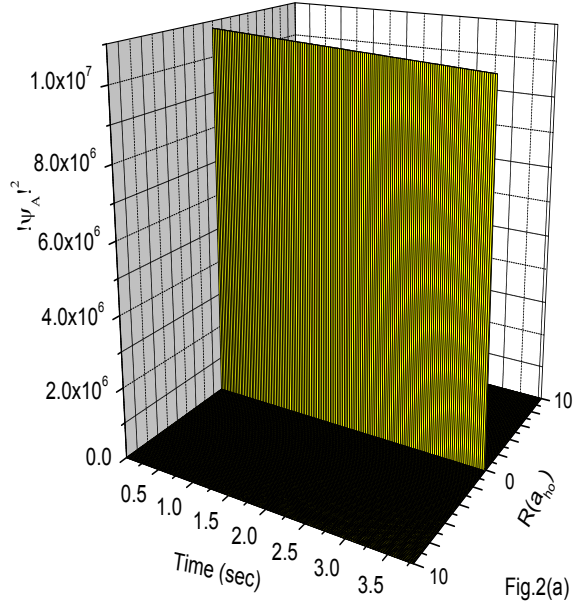}
\includegraphics[width=3.5in,height=3.5in,angle=0]{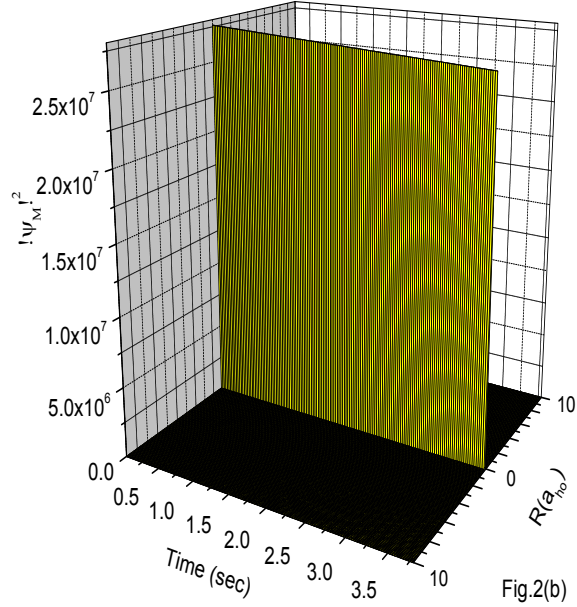}
\caption {Fig 2. Dynamical evolution of (a) atomic and (b) molecular solitons in a hybrid atom-molecular coupled BEC system of $^{85}Rb$ atom (through Magnetic Feshbach Resonance (MFR)) with trap on for ~3.75 secs in an effective one dimensional trap of frequency $\omega = 12.69 Hz$. $N=17100$, $\Gamma_1= 1.629\times 10^{-13}$ $m^3 Hz$ and $\Gamma_2 = 304.4 \times 10^{12} Hz$, $a = 3.016308 \times 10^{-8} m $ and other parameters are the same as given at the end of section 'Theory'. Here $R(a_{ho})$ is the radial distance from the centre of the trap. }
\end{figure}
\begin{figure}[!h]
\includegraphics[width=3.5in,height=3.5in,angle=0]{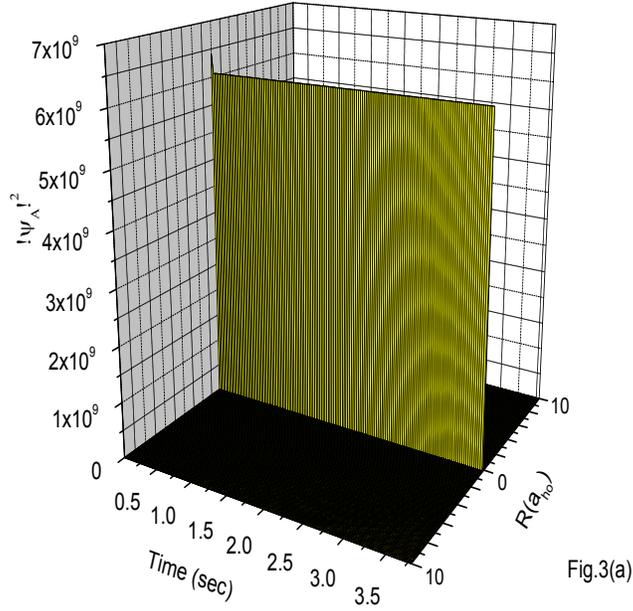}
\includegraphics[width=3.5in,height=3.5in,angle=0]{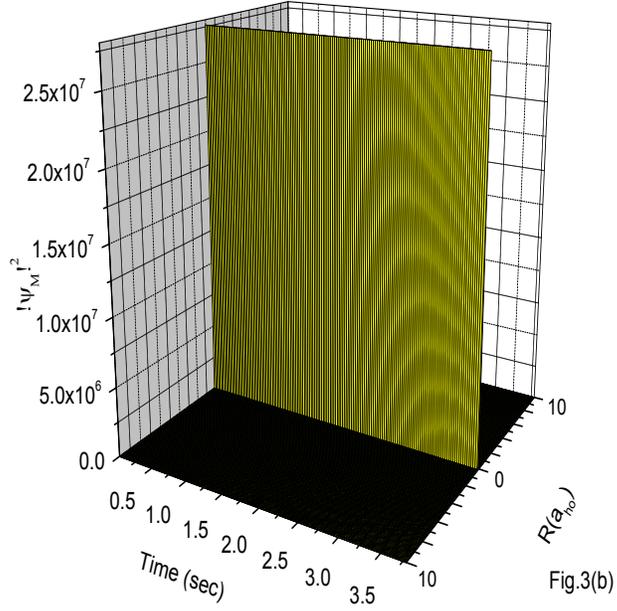}
\caption {Fig 3. Dynamical evolution of (a) atomic and (b) molecular solitons in a hybrid atom-molecular coupled BEC system of $^{85}Rb$ atom (through MFR) with trap on for ~3.75 secs. $\Gamma_1=1.629 \times 10^{-17} m^3 Hz$ and $\Gamma_2 = 304.4 \times 10^{14} Hz$
and other parameters used are the same as given in Fig.2}
\end{figure}
\begin{figure}[!h]
\includegraphics[width=3.5in,height=3.5in,angle=0]{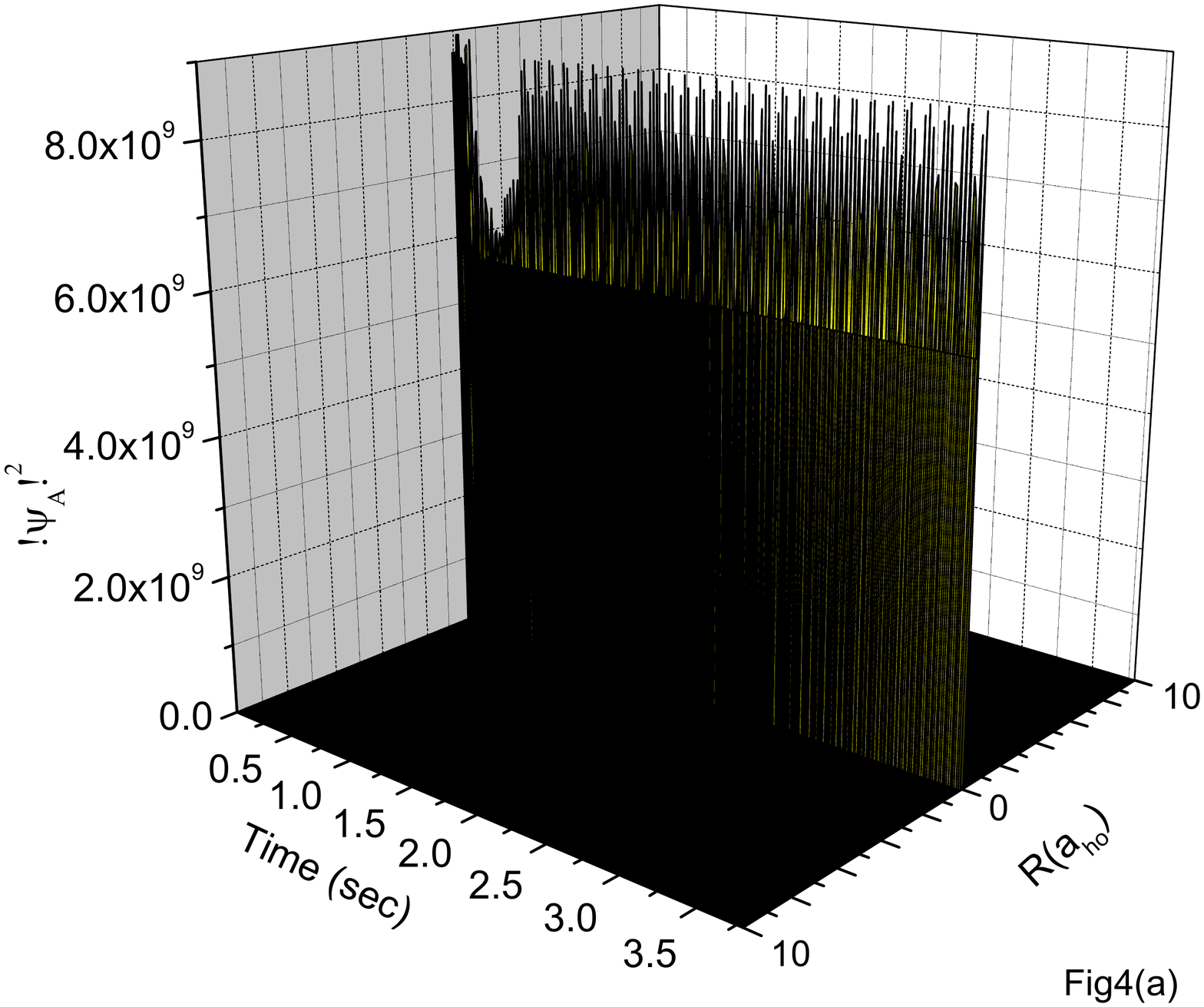}
\includegraphics[width=3.5in,height=3.5in,angle=0]{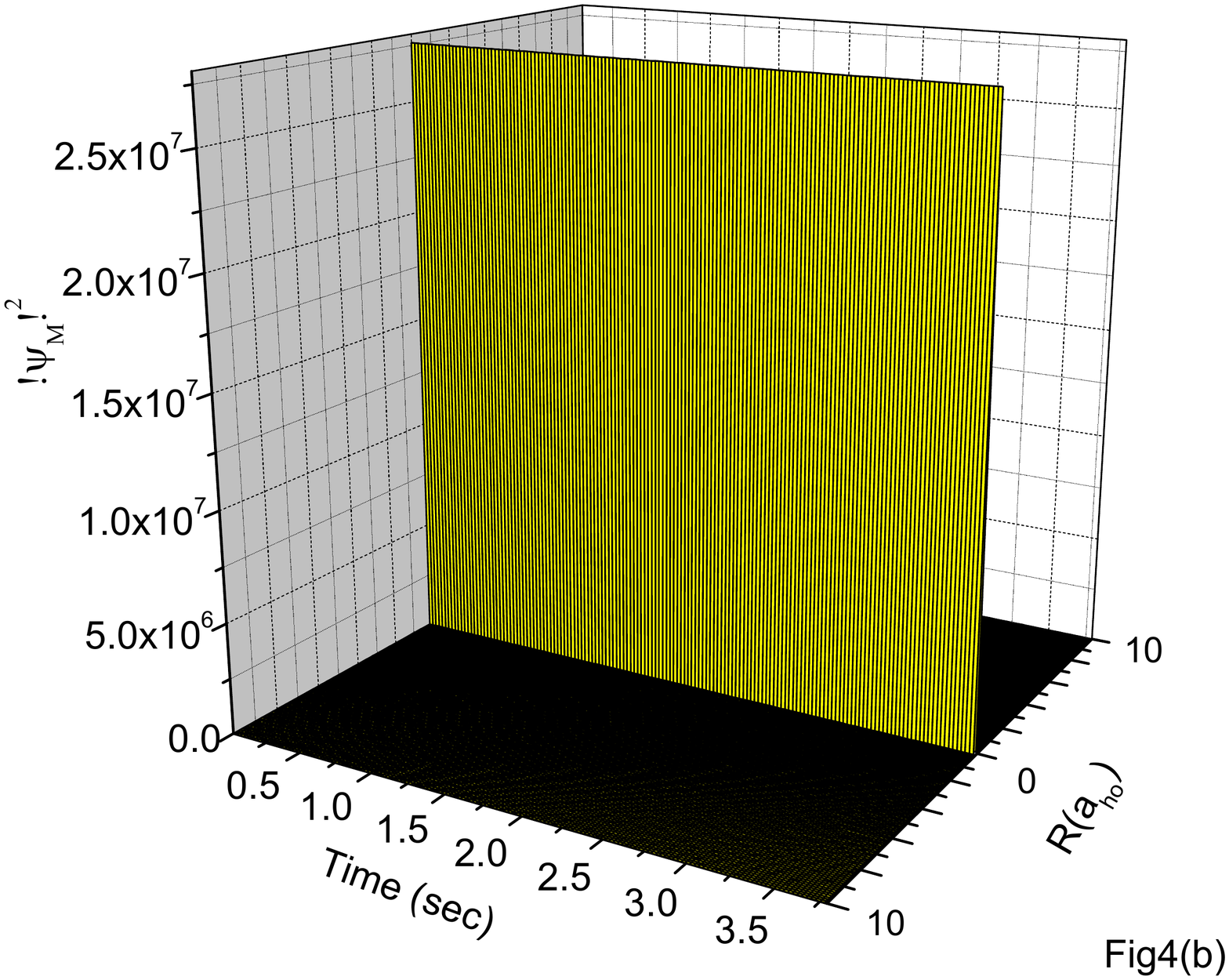}
\caption {Fig 4. Dynamical evolution of (a) atomic and (b) molecular solitons in a hybrid atom-molecular coupled BEC system of $^{85}Rb$ atom (through MFR) with trap on for 3.75 secs, $\Gamma_1=1.629 \times 10^{-21} m^3 Hz$ and $\Gamma_2 = 304.4 \times 10^{14} Hz$.
Other parameters used are the same as given in Fig.2}
\end{figure}
\end{document}